\documentclass[aps,prl,twocolumn,amsmath,amssymb,showpacs,superscriptaddress,notitlepage]{revtex4-1}
\usepackage{amsmath,amssymb,amsfonts,bm}
\usepackage{graphicx}
\usepackage{epstopdf}
\usepackage{dcolumn}
\usepackage{mathrsfs}
\usepackage{bbold}
\usepackage{dsfont}
\usepackage{dcolumn}
\usepackage[colorlinks=true,linkcolor=blue,citecolor=blue, urlcolor=blue,bookmarks=false]{hyperref}

\begin{document}

\title{Using nonlocal surface transport to identify the axion insulator}

\author{Rui Chen}
\affiliation{Shenzhen Institute for Quantum Science and Engineering and Department of Physics, Southern University of Science and Technology (SUSTech), Shenzhen 518055, China}
\affiliation{School of Physics, Southeast University, Nanjing 211189, China}

\author{Shuai Li}
\affiliation{Shenzhen Institute for Quantum Science and Engineering and Department of Physics, Southern University of Science and Technology (SUSTech), Shenzhen 518055, China}

\author{Hai-Peng Sun}
\affiliation{Shenzhen Institute for Quantum Science and Engineering and Department of Physics, Southern University of Science and Technology (SUSTech), Shenzhen 518055, China}

\author{Qihang Liu}
\affiliation{Shenzhen Institute for Quantum Science and Engineering and Department of Physics, Southern University of Science and Technology (SUSTech), Shenzhen 518055, China}

\author{Yue Zhao}
\affiliation{Shenzhen Institute for Quantum Science and Engineering and Department of Physics, Southern University of Science and Technology (SUSTech), Shenzhen 518055, China}
\affiliation{Shenzhen Key Laboratory of Quantum Science and Engineering, Shenzhen 518055, China}

\author{Hai-Zhou Lu}
\email{Corresponding author: luhz@sustech.edu.cn}
\affiliation{Shenzhen Institute for Quantum Science and Engineering and Department of Physics, Southern University of Science and Technology (SUSTech), Shenzhen 518055, China}
\affiliation{Shenzhen Key Laboratory of Quantum Science and Engineering, Shenzhen 518055, China}

\author{X. C. Xie}
\affiliation{International Center for Quantum Materials, School of Physics, Peking University, Beijing 100871, China}
\affiliation{Beijing Academy of Quantum Information Sciences, Beijing 100193, China}
\affiliation{CAS Center for Excellence in Topological Quantum Computation, University of Chinese Academy of Sciences, Beijing 100190, China}

\begin{abstract}
The axion is a hypothetical but experimentally undetected particle. Recently, the antiferromagnetic topological insulator MnBi$_2$Te$_4$ has been predicted to host the axion insulator, but the experimental evidence remains elusive. Specifically,
the axion insulator is believed to carry ``half-quantized" chiral currents running antiparallel on its top and bottom surfaces. However, it is challenging to measure precisely the half-quantization. Here, we propose a nonlocal surface transport device, in which the axion insulator can be distinguished from normal insulators without a precise measurement of the half-quantization. More importantly, we show that the nonlocal surface transport, as a qualitative measurement, is robust in realistic situations when the gapless side surfaces and disorder come to play. Moreover, thick electrodes can be used in the device of MnBi$_2$Te$_4$ thick films, enhancing the feasibility of the surface measurements. This proposal will be insightful for the search of the axion insulator and axion in topological matter.
\end{abstract}
\maketitle
\begin{figure}[h!]
\centering
\includegraphics[width=0.95\columnwidth]{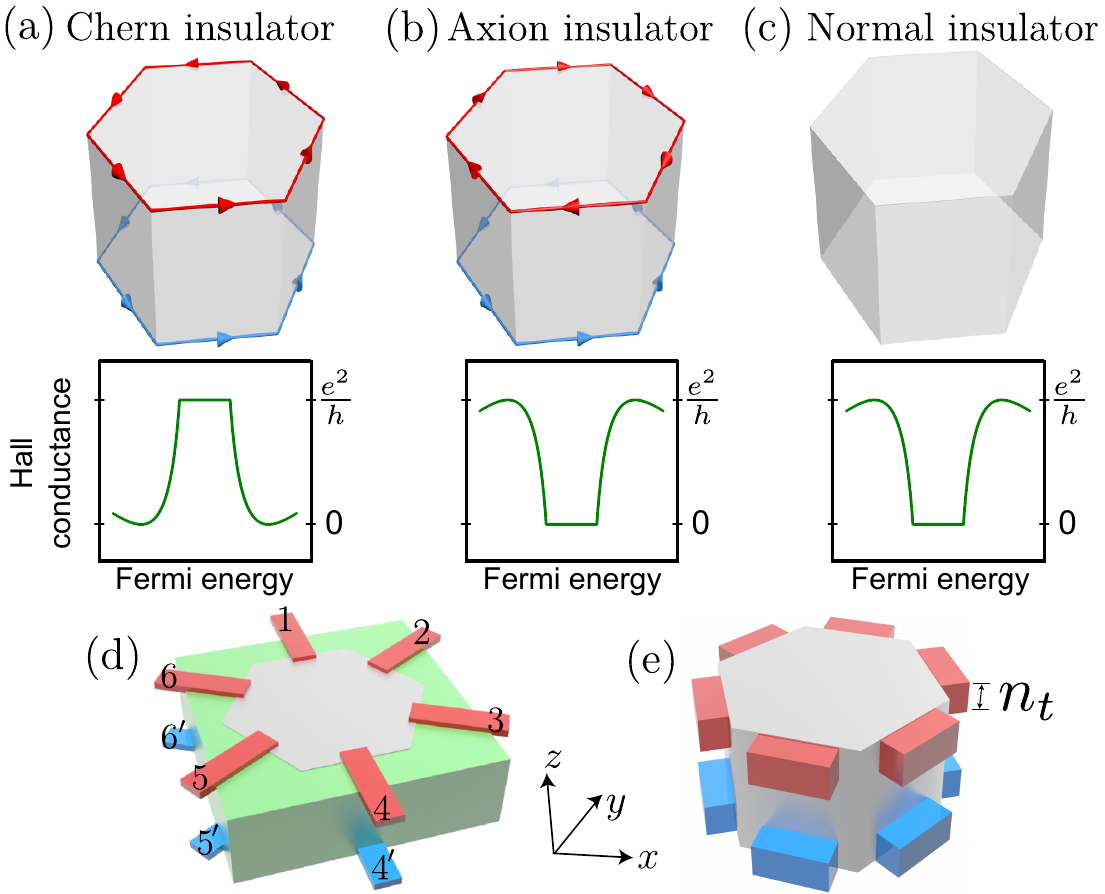}
\caption{For the antiferromagnetic topological insulator MnBi$_2$Te$_4$, (a) odd septuple layers host a Chern insulator with parallel ``half-quantized" chiral currents on the top and bottom surfaces, and (b) even septuple layers may support the axion insulator with opposite chiral currents. (c) The normal insulator has no chiral currents. (d-e) The proposed device for nonlocal surface measurements. Red, blue, green, and grey correspond to the top and bottom electrodes, insulating shield, and antiferromagnetic topological insulator, respectively. $n_t$ is the thickness of the thick electrodes. To distinguish the phases in the MnBi$_2$Te$_4$ films, one first measures the Hall conductance $\sigma_{xy}$. If $\sigma_{xy}=\pm e^2/h$, it is (a) a Chern insulator. If $\sigma_{xy}=0$, it could be either (b) an axion insulator or (c) a normal insulator. The axion insulator can be further distinguished from the normal insulator, because the axion insulator has different nonlocal signals on the top and bottom surfaces while the normal insulator is supposed to have no nonlocal signal. We also check that the topological insulator and Dirac semimetal have no nonlocal signal because of time-reversal symmetry and the Weyl semimetal has the same nonlocal signals on the top and bottom surfaces, all can be distinguished from the axion insulator.}
\label{fig_surfaceT}
\end{figure}

{\color{blue}\emph{Introduction}.}--
The axion is an elementary particle postulated to resolve the strong CP problem in quantum chromodynamics, but remains invisible in experiments~\cite{Peccei77prl}. In recent years, the axion insulator in condensed matter physics has attracted great attention because it shares the axionic electromagnetic response~\cite{Mogi17nm,Mogi17sa,Xiao18prl,Varnava18prb,Liu20prb,Xu19prl}, which modifies Maxwell's equations and may lead to a ``half-quantized" surface Hall conductance~\cite{Essin09prl,Mong10prb,Niu20prl,Xu14np} or the topological magnetoelectric effect~\cite{Wang15prbrc,Morimoto15prb,Qi09sci,Maciejko10prl,Tse10prl,Yu19prb}. Unfortunately, these signatures require highly demanding precisions of measurement and device fabrication, so the experimental attempts to identify the axion insulator have yet to be successful~\cite{Liu20prb,Varnava18prb}~(see also Sec.~IX of~\cite{Supp}). Recently, in the experiments of the first antiferromagnetic topological insulator MnBi$_2$Te$_4$~\cite{Chang13sci,Yu10sci,Zhang19prb,Wang15ps,Ding20prb,
Zhang19prl,Li19sa,Otrokov19prl,
Otrokov19nat,Deng20sci,Liu20nm,
Gong19cpl,Zhang20prl,Li19prb,Shi19prb,
Chen19prx,Hao19prx,Sun19prl,Jo19arXiv}, the signatures for the quantized anomalous Hall effect and axion insulator have been reported~\cite{Deng20sci,Liu20nm}.
Specifically, the axion (Chern) insulator has antiparallel (parallel) chiral currents on the top and bottom surfaces. Each current is responsible for the $e^2/2h$ Hall conductance. The top and bottom chiral currents combine to yield a zero ($e^2/h$) Hall conductance as a signature for the axion (Chern) insulator [Figs.~\ref{fig_surfaceT}(a)-(b)]. The Chern and axion insulators may be distinguished by optical measurements~\cite{Pozo2019PRL}.
However, normal insulators also have a zero Hall conductance [Fig.~\ref{fig_surfaceT}(c)]. Therefore, seeking signatures for the axion insulator is still an open problem.

In this Letter, we propose that the axion insulator can be distinguished from normal insulators by measuring nonlocal surface resistances in a device of the antiferromagnetic topological insulator MnBi$_2$Te$_4$ [Fig.~\ref{fig_surfaceT}(d)]. MnBi$_2$Te$_4$ is formed by stacked septuple layers with out-of-plane magnetization. The neighbouring septuple layers have opposite magnetizations. For even (odd) layers, the chiral currents on the top and bottom surfaces propagate along the opposite (same) directions [Figs.~\ref{fig_surfaceT}(a)-(b)], forming an axion insulator (a thick Chern insulator). Because of the chiral currents, the axion insulator shows distinct nonlocal surface transport, compared to those in the Chern insulator and normal insulator. We numerically calculate the nonlocal surface transport of MnBi$_2$Te$_4$ with different septuple layers by using a lattice model and realistic parameters. More importantly, we show that the nonlocal surface transport, as a qualitative measurement, is robust against disorder andside-surface transport. Moreover, we show that thick films of MnBi$_2$Te$_4$ allow thick electrodes to measure the surface transport [Fig.~\ref{fig_surfaceT}(e)]. Therefore, the nonlocal surface transport provides a feasible approach to identify the axion insulator. Our results will be helpful for the ongoing and future search for the axion insulator.

\begin{figure}[tpb]
\centering
\includegraphics[width=0.9\columnwidth]{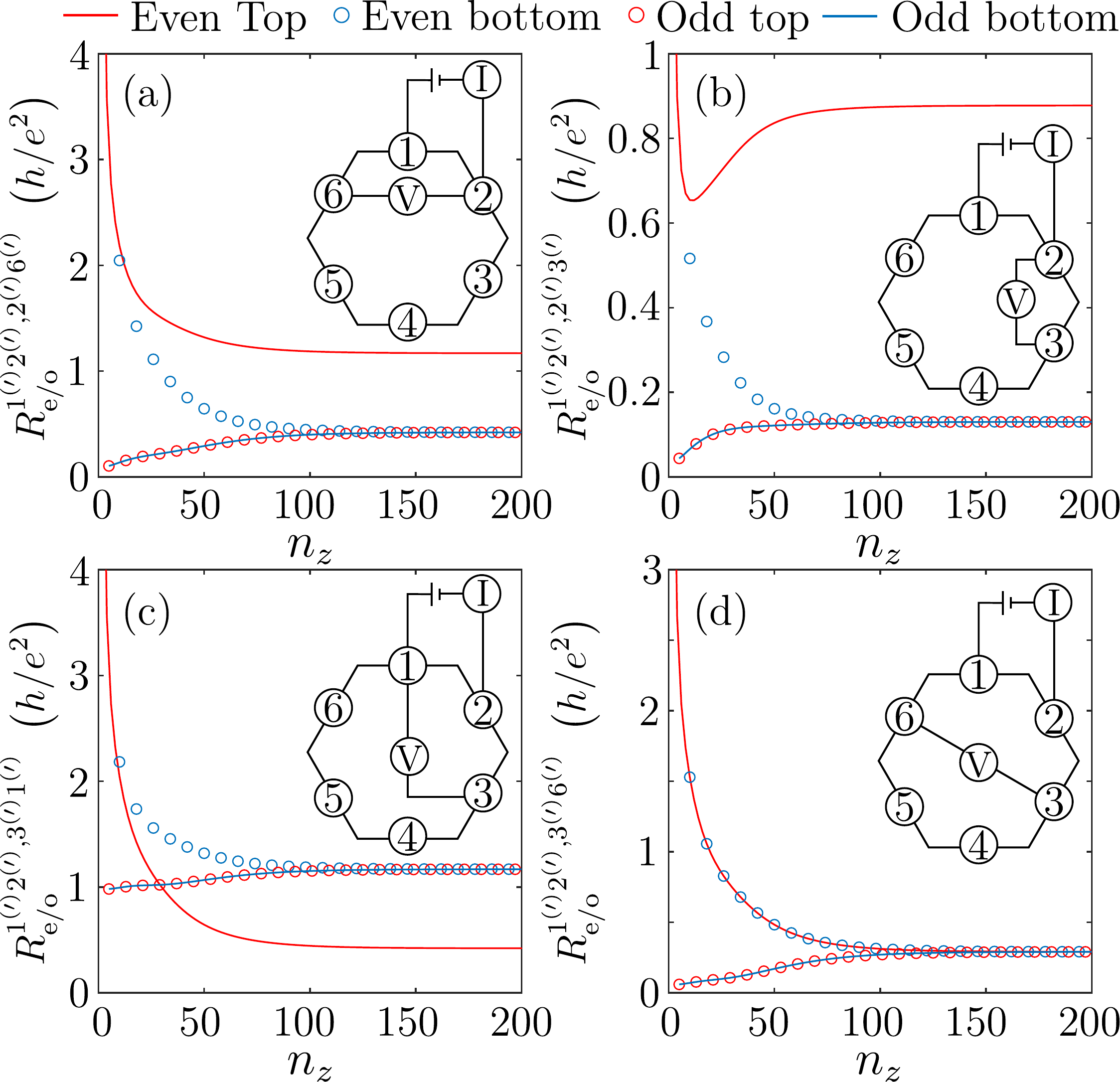}
\caption{The numerically calculated nonlocal surface resistances defined as $R=V/I$ in the insets.
This figure presents only ideal cases for demonstration. The side surface, disorder, and thick electrodes are considered in Figs. \ref{fig_vary_Ef}, \ref{fig_vary_disorder}, and \ref{fig_thickerleads}, respectively. The red line and blue circle (red circle and blue line) correspond to the top and bottom surface of an even (odd) $n_z$, respectively.}
\label{fig_illustration}
\end{figure}

{\color{blue}\emph{Nonlocal surface transport}.}--
The axion insulator is characterized by the opposite ``half-quantized" chiral currents [Fig.~\ref{fig_surfaceT}(b)] due to its axion electromagnetic response. Detecting the chiral current can help to identify the axion insulator. Compared to the ultrathin films of magnetically-doped Bi$_{2}$Se$_3$-like materials, bulk crystals of intrinsic antiferromagnetic insulator MnBi$_2$Te$_4$ can be used directly for measurements, and allow electrodes to probe mostly the top or bottom surface, respectively, as shown in Fig.~\ref{fig_surfaceT}(d).
Usually, a direct probe of the chiral currents is prohibited by the side surfaces, which have metallic surface states that bury the signals of the chiral currents. To solve this problem, we introduce the nonlocal surface measurements. The metallic states on the side surfaces are not chiral, that is, they go opposite directions with the same weight, thus do not play a significant role in the nonlocal transport compared to the chiral currents. Specifically, the top (bottom) surface of the device has six electrodes, $1$-$6$ ($1'$-$6'$)~[Fig.~\ref{fig_surfaceT}(d)].
As shown in Fig.~\ref{fig_illustration}, a current $I$ is applied between electrodes 1 and 2 ($1^{\prime}$ and $2^{\prime}$) on the top (bottom) surface, and a nonlocal voltage $V$ is measured between other two electrodes on the same surface, to define a nonlocal resistance $R=V/I$~\cite{Roth09sci}.

For a given surface, the nonlocal resistances can be analytically~\cite{Datta1997} found for the 4 configurations in Fig.~\ref{fig_illustration} as

\begin{eqnarray}\label{resistance}
R^{12,26} &=&\frac{h}{e^{2}}\frac{n_{L}F_{2}}{F_{4}+G_{2}},R^{12,31}=\frac{h%
}{e^{2}}\frac{n_{R}F_{2}}{F_{4}+G_{2}},  \notag
\\
R^{12,23} &=&\frac{h}{e^{2}}\frac{n_{L}^{4}}{F_{1}\left( F_{4}+G_{2}\right) },
R^{12,36}=\frac{h}{e^{2}}\frac{n_{L}n_{R}}{F_{3}}
\end{eqnarray}%
with $F_{i}=n_{L}^{i}+n_{R}^{i}$ and $G_{i}=n_{L}^{i}n_{R}^{i}$. Here, $n_L$ and $n_R$ determine the matrix elements the transmission probability matrices (details and more configurations can be found in Sec. SII of \cite{Supp}).

Before considering realistic situations with side surfaces, disorder, and thick electrodes, we first use ideal cases for illustration. For odd layers, the half-quantized chiral currents on the top and bottom surfaces are parallel, i.e., $n_L=0$ and $n_R=1/2$ for both top and bottom surfaces. For even layers, the half-quantized chiral currents are opposite, i.e., $n_L=0$ and $n_R=1/2$ on the top and
$n_L=1/2$ and $n_R=0$ on the bottom surface.
Table~\ref{tab1} shows the the ideal nonlocal resistances for the configurations in Fig.~\ref{fig_illustration}.
For odd layers, the two surfaces always have the same nonlocal resistance. For even layers, the nonlocal resistances on the two surfaces have different values for most of the cases, except for the case with $p,q=3,6$.
Therefore, the axion insulator in even layers and Chern insulator in odd layers would show distinct nonlocal surface transport properties. More importantly, the normal insulator has no such nonlocal resistance because it has no chiral current (see Fig.~\ref{fig_surfaceT}(c) and Sec.~SVI of \cite{Supp}). Therefore,  the unique nonlocal resistance can be used to distinguish the axion insulator from normal insulators.
Above are the ideal cases. In real materials with side surfaces, disorder, and thick electrodes, the chiral currents $n_L$ and $n_R$ are not perfectly half-quantized. Below we will verify the above proposal by using a realistic model and simulations.


\renewcommand\arraystretch{1.5}
\begin{table}[t]
\begin{ruledtabular}
\caption{The ideal resistances $R_{\text{e/o}}^{1^{(\prime)}2^{(\prime)},p^{(\prime)}q^{(\prime)}}$ obtained from the analytical Eq.~(\ref{resistance}) for the configurations in Fig.~\ref{fig_illustration}.}
\begin{tabular}{cccccc}
$p,q$ & $2,6$ & $3,1$& $2,3$& $3,6$ \\
\hline
$R_{\text{e}}^{12,pq}\left( h/e^{2}\right) $ & 2 & 0 & 2 & 0 \\
\hline
$R_{\text{e}}^{1^{\prime }2^{\prime },p^{\prime }q^{\prime }}\left( h/e^{2}\right) $
& 0 & 2 & 0 & 0  \\ \hline
$R_{\text{o}}^{12,pq}\left( h/e^{2}\right) $ & 0 & 2 & 0 & 0  \\
\hline
$R_{\text{o}}^{1^{\prime }2^{\prime },p^{\prime }q^{\prime }}\left( h/e^{2}\right) $
& 0 & 2 & 0 & 0  \\
\end{tabular}
\label{tab1}
\end{ruledtabular}
\end{table}
\renewcommand\arraystretch{1}

{\color{blue}\emph{Model}.}--
For numerical calculations, we use an effective model for MnBi$_2$Te$_4$ regularized on a stacked hexagonal lattice \cite{Zhang20prl}
\begin{equation}
H=%
\begin{pmatrix}
h +\mathbf{M}_{A}\cdot s\otimes \sigma _{0}& h_{AB} \\
h_{AB}^{\dag } & h+\mathbf{M}_{B}\cdot s\otimes \sigma _{0}
\end{pmatrix}%
,
\end{equation}%
with the intra-layer part
$h $ = $[ \tilde{C}$ - $(4/3)C_{2} $ $( \cos k_{1}$ + $\cos k_{2} $ + $\cos
k_{3} ) ] $ $ I_{4} $
+$(v/3) $ $ ( 2\sin k_{1} $ + $\sin k_{2}+\sin k_{3}) $ $\Gamma _{1} $
+$(v/\sqrt{3}) $ $ ( \sin k_{2}-\sin k_{3}) $ $\Gamma _{2}$
+$[ \tilde{M}$ - $(4/3)M_{2} ( \cos k_{1}+\cos k_{2}+\cos
k_{3}) ]$$ \Gamma _{4}$, the inter-layer part
$h_{AB}$ $=-2C_{1}\cos k_{z}I_{4}$ + $v_{z}\sin k_{z}\Gamma _{3}-2M_{1}\cos k_{z}$$ \Gamma _{4}$.
The lattice constants $\mathbf{a}_{1}=\left( 1,0,0\right), \mathbf{a}_{2}=\left( 1,\sqrt{3},0\right)/2, \mathbf{a}_{3}=\left(
0,0,1\right)$. The wave vectors $k_{1}=k_{x}$, $k_{2}=\left( k_{x}+\sqrt{3}k_{y}\right)/2 \,$, $k_{3}=k_{1}-k_{2}=\left(
k_{x}-\sqrt{3}k_{y}\right)/2 $.
$I$ is the identity matrix, $\Gamma _{i}=s_{i}\otimes \sigma _{1}
$ for $i=1,2,3$,
$\Gamma_{4}=s_{0}\otimes \sigma _{3}$, $\tilde{M}=M_0+2M_1+4M_2$, and $\tilde{C}=C_0+2C_1+4C_2$. $s_{i}$ and $\sigma _{i}$ are the Pauli
matrices. $\mathbf{M}_{k}=m\left( \cos
\phi _{k}\sin \theta _{k},\sin \phi _{k}\sin \theta _{k},\cos \theta
_{k}\right) $ with angles $\phi _{k}$ and $\theta _{k}$ for $k=A,B$.
The antiferromagnetic
order is described by $\left( \phi _{A},\theta _{A}\right) =\left(
0,0\right) $ and $\left( \phi _{B},\theta _{B}\right) =\left( 0,\pi \right) $.
For a realistic simulation, we use the parameters transformed from those in the $k\cdot p$ model ~\cite{Zhang19prl,Supp}, $C_{0}=C_{1}=C_{2}=0$, $M_{0}=-0.1165$ eV, $M_{1}=11.9048$ eV\AA$^2$, $M_{2}=9.4048$ eV\AA$^2$, $v_{z}=2.7023$ eV\AA, $v=3.1964$ eV\AA, and $m=0.1$ eV, which describes the A-type antiferromagnetic topological insulator in MnBi$_2$Te$_4$.

\begin{figure}[htbp]
\centering
\includegraphics[width=0.9\columnwidth]{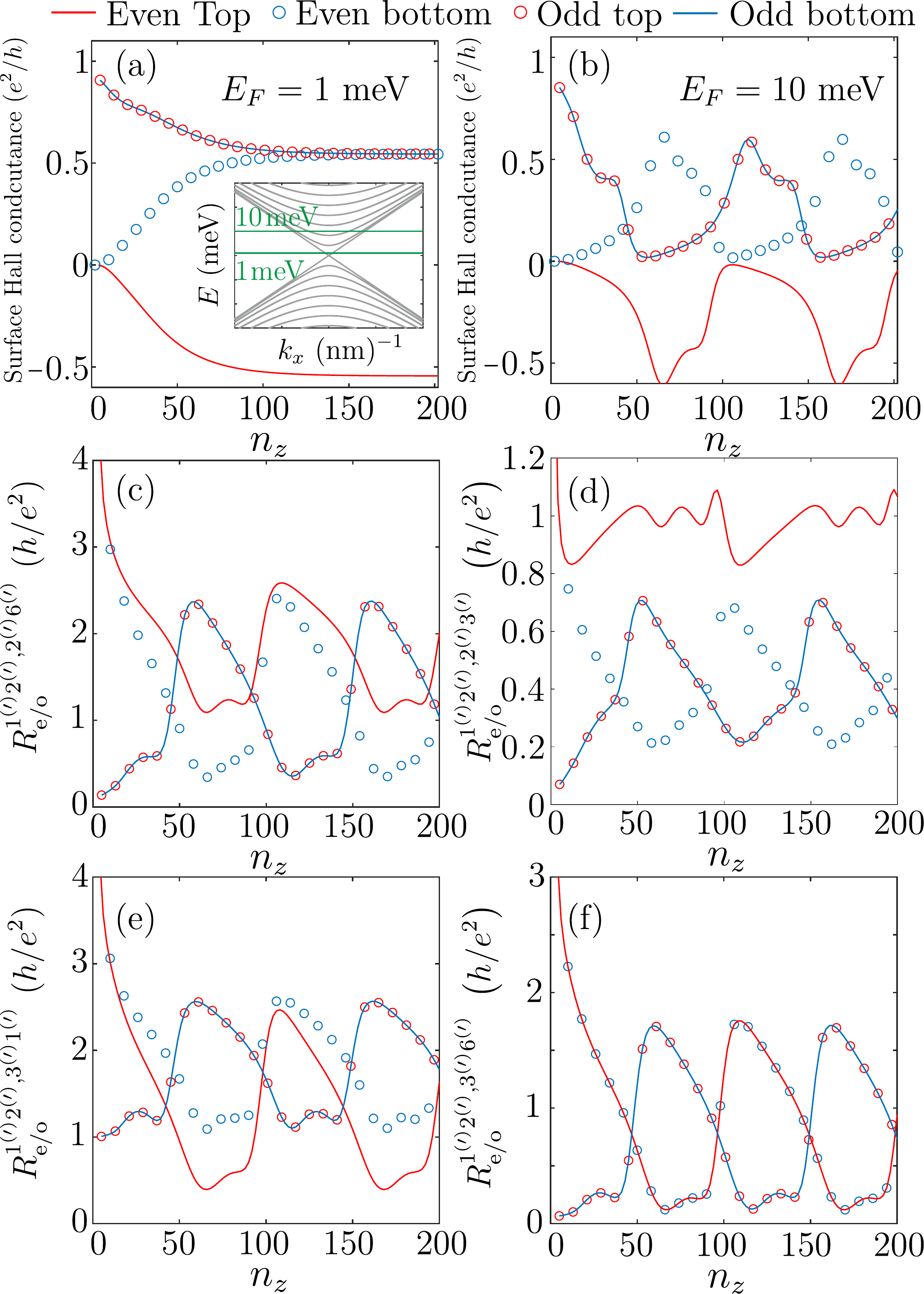}
\caption{(a-b) The surface Hall conductance on the top and bottom surface as functions of the number of septuple layers $n_z$ with the Fermi energy $E_F=1$ meV and $E_F=10$ meV, respectively. The inset in (a) shows the subbands of the side surface in the $x$-$z$ plane. The green lines indicate $E_F=1$ and $10$ meV. (c-f) The numerically calculated nonlocal surface resistances
for $E_F=10$ meV. The red lines and blue circles (red circles and blue lines) correspond to the top and bottom surface of an even (odd) $n_z$, respectively. See the results for other configurations in Sec.~SIV  of \cite{Supp}.
}
\label{fig_vary_Ef}
\end{figure}

The model describes a topological insulator with two gapped surface states on the top and bottom surfaces (Fig. S1(b) of \cite{Supp}), due to breaking of the combined antiferromagnetic time-reversal symmetry $\Theta_M$.
The gapped surfaces behave like ``marginal" 2D Chern insulators, characterized by ``half-quantized" chiral currents (Fig.~\ref{fig_surfaceT}), in a sense that their Chern numbers are half-quantized as $\pm 1/2$, or in other words the Hall conductance
$\sigma_{xy}= \pm e^2/2h$ [see, e.g., Fig.~\ref{fig_vary_Ef} (a) as $n_z\rightarrow 200$]. The sign $\pm$ depends on the magnetization, which reverses from one septuple layer to another in MnBi$_2$Te$_4$~\cite{Mong10prb,Li19sa}. For odd (even) layers, the top and bottom surfaces have the same (opposite) magnetization, so the half-quantized surface Hall conductances of the top and bottom surfaces add up (cancel) to give a Chern insulator characterized by $C=1$ (axion insulator with $C=0$)~\cite{Zhang19prl,Li19sa,Otrokov19prl,
Otrokov19nat,Deng20sci,Liu20nm,
Gong19cpl,Zhang20prl,Li19prb,Shi19prb,
Chen19prx,Hao19prx} (see also Fig. S1(c) of \cite{Supp}).
The chiral currents, which are responsible for the half-quantized surface Hall conductance, can be numerically verified using first-principles calculations for MnBi$_2$Te$_4$ \cite{Gu20arXiv}.

{\color{blue}\emph{Calculation of the nonlocal surface transport}.}--We calculate the nonlocal surface transport~\cite{Chu11prb} by using the Landauer-B\"uttiker-Fisher-Lee formula
\cite{Landauer70pm,Buttiker88prb,
Fisher81prb} and the recursive Green's
function method \cite{Mackinnon85zpb,Metalidis05prb}.
At zero temperature, the current flowing into electrode $p$ is given by
\begin{eqnarray}
I_p &=& \frac{e^2}{h} \sum_{q\neq p} T_{pq}(E_F)(V_p-V_q),
\end{eqnarray}
where $V_{p}$ is the voltage at electrode $p$, $T_{pq}$ is the transmission coefficient from electrode $q$ to $p$.
For the present case, there are 6 electrodes for each of top and bottom surfaces, so $T_{pq}$ is a 6$\times$6 matrix (more details can be found in Sec. SII and Sec. SVI of \cite{Supp}).

Figure \ref{fig_illustration} shows the numerically calculated nonlocal resistances for 4 configurations as functions of film thickness (number of septuple layers $n_z$). In odd layers, the resistances have the same values for the top and bottom surfaces. In even layers, the resistances have distinct values for the two surfaces, except for those in Fig.~\ref{fig_illustration}(d). For thick films in Fig.~\ref{fig_illustration}, there are deviations from the analytic values in Tab.~\ref{tab1}. This is due to the gapless side surface states~\cite{Chu11prb}, and is one of the reasons why it is hard to measure the half-quantized Hall effect as the signature for the axion insulator. Later, we will show the nonlocal surface transport, as a qualitative measurement, is robust against not only the side-surface transport but also disorder and thick electrodes.

\begin{figure}[tpb]
\centering
\includegraphics[width=0.9\columnwidth]{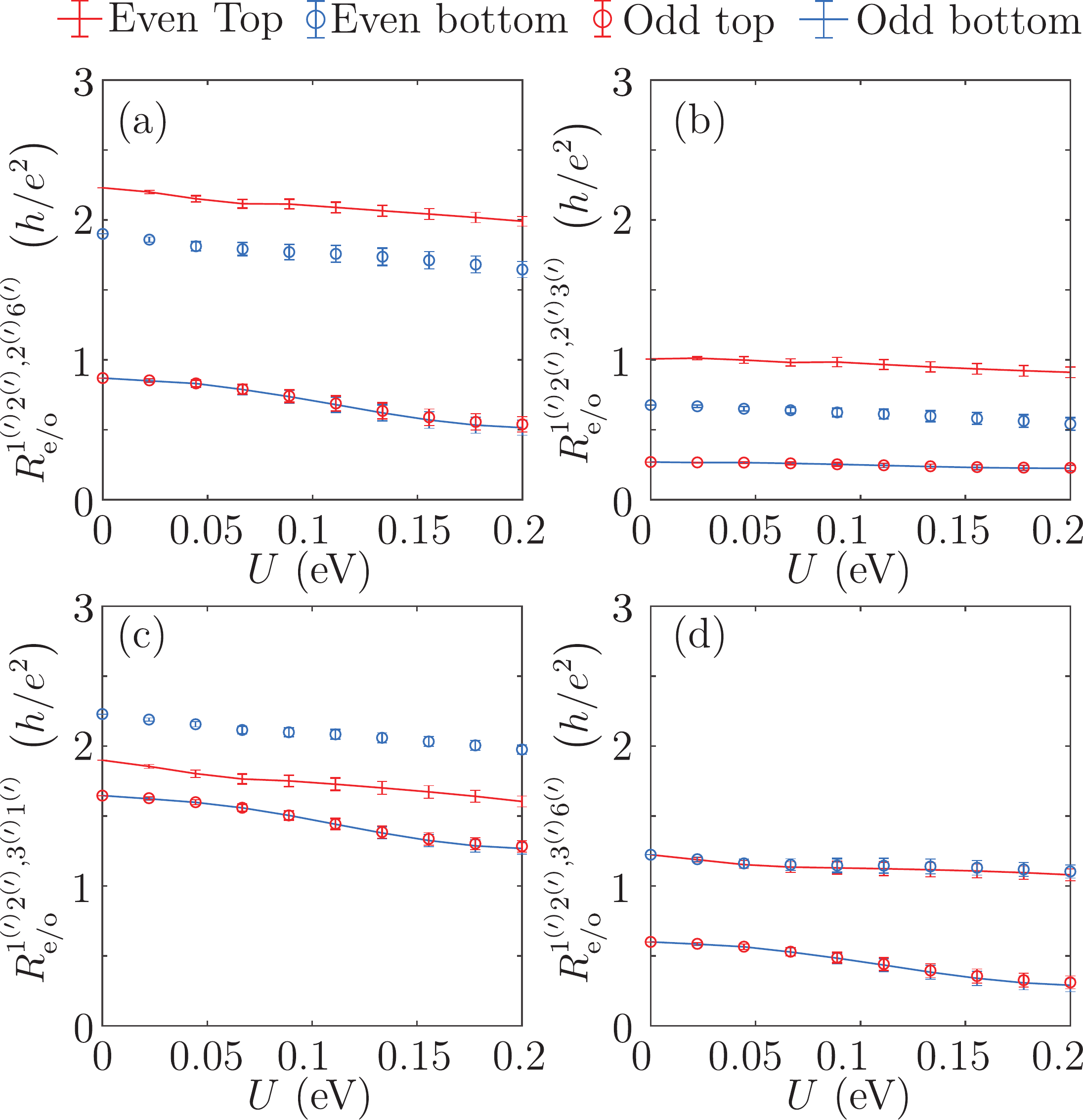}
\caption{The nonlocal surface resistances as functions of the disorder strength $U$ with $E_F=10$ meV. The error bars show the standard deviation of the conductance for 100 samples.
Red line and blue circle (red circle and blue line) correspond to the top and bottom surface of an even (odd) $n_z$, respectively. See results for other configurations in Sec.~SIV of \cite{Supp}.}
\label{fig_vary_disorder}
\end{figure}

{\color{blue}\emph{Side-surface effects}}.--
Figures~\ref{fig_vary_Ef}(a-b) show the numerical results of the surface Hall conductance for different $E_F$, which can be found from the left-chiral and
right-chiral currents, e.g., $e^2/h(T^{12}_{\text{e}}-T^{21}_{\text{e}})$ for the top surface in even layers.
For the ideal case, the top and bottom surface Hall conductances are half-quantized ($\pm e^2/2h$), and reduce
to 0 and $e^2/h$ as $n_z\rightarrow 0$ for even and odd layers, respectively, as a result of the top-bottom hybridization~\cite{Zhou08prl,Lu10prb,Linder09prb,ZhangY10np,Chen16cpb,Liu10prb,Imura12prb,Takane16jpsj,Chen17prb}. A zoom-in of Figure~\ref{fig_vary_Ef}(a) shows that the surface Hall conductances are not precisely half-quantized due to the coupling of the side surface states when the Fermi energy is away from the Dirac point. This shows that a direct measurement of the half-quantized Hall effect is challenging, in accordance with the previous studies~\cite{Konig14prb}. Figure~\ref{fig_vary_Ef}(b) shows that the side surface states turn the surface Hall conductance into an oscillation with $n_z$ at higher Fermi energies ($E_F=10$ meV), as a result of the confinement-induced subbands of the side surface states [inset of Fig.~\ref{fig_vary_Ef}(a)]. Similar oscillations also present in the nonlocal surface transport in Figs.~\ref{fig_vary_Ef}(c-f).


{\color{blue}\emph{Disorder effects}}. -- Now, we show that the nonlocal surface transport is robust against disorder.
We introduce the Anderson-type disorder to the central scattering region with $\Delta H=\sum_{i}{W_{i}\sigma_{0}c_{i }^{\dag
}c_{i }}$, where $W_{i}$ is uniformly distributed within $\left[-U/2,U/2\right]$, with $U$ being the disorder strength. In Fig.~\ref{fig_vary_disorder}, we show the nonlocal surface resistance as functions of the disorder strength $U$ for $n_z=200$ and with the side surfaces ($E_F=10$ meV).
Even for strong disorder ($U=0.2$ eV, about the size of the bulk band gap), the nonlocal resistances remain at the same order of magnitude.

\begin{figure}[tpb]
\centering
\includegraphics[width=0.9\columnwidth]{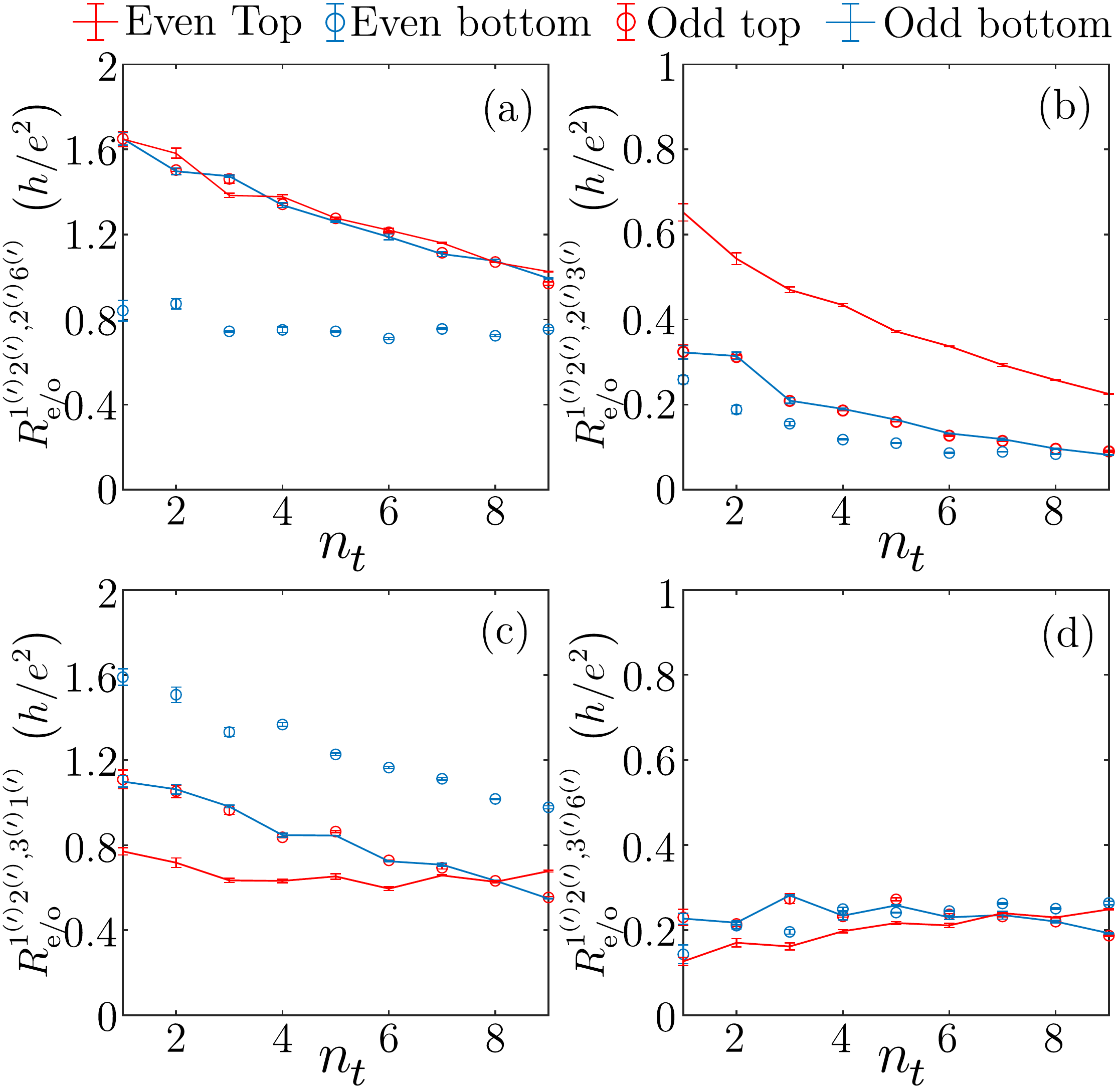}
\caption{Schematic illustration of the 3D device with thick electrodes. Here the thickness of the electrodes is $n_t$. (a-d) show the numerically calculated nonlocal surface resistances as functions of $n_t$ with $E_F=10$ meV and $U=100$ meV. The error bars show the standard deviation of the conductance for 100 samples.
Red line and blue circle (red circle and blue line) correspond to the top and bottom surface of an even (odd) $n_z=100\text{ }(101)$, respectively. See more details in Sec. SVIII of \cite{Supp}.}
\label{fig_thickerleads}
\end{figure}

{\color{blue}\emph{Thick electrodes}}. -- Above, we assume that the electrodes are only attached to the topmost and bottommost septuple layers of the device.
Figure \ref{fig_thickerleads} shows the numerical results for thick electrodes [Fig.~\ref{fig_surfaceT}(e)]. We consider a realistic situation in which the side surface states ($E_F=10$ meV) and disorder ($U=100$ meV) are also taken into account. The nonlocal resistances drop with increasing electrode thickness ($n_t$), but stay at the same order of magnitude up to 10 septuples (about 14 nm, thicker than the thinnest metal electrode that can be fabricated). The chiral nature of the surface transport for even and odd layers are still stable with increasing the electrode thickness, showing the topological nature of the chiral currents.
%

{\color{blue}\emph{Discussion and experimental realization}}.--
In this work, we focus on distinguishing the axion insulator and normal insulator, by using the nonlocal signal. If and only if the system is an axion insulator, the nonlocal resistances are significantly different on the top and bottom surfaces. Furthermore, we also calculated the surface transport coefficients for those systems without the chiral hinge currents, including the topological insulator, Weyl semimetal, and Dirac  semimetal (see Sec.~SVI of \cite{Supp}). Our calculations show that, the nonlocal signals in the axion insulator are absent in other systems without the chiral hinge currents. Moreover, the propagation directions of surface chiral states can be identified if the nonlocal resistances can be measured by using properly chosen combinations of the measurement electrodes, e.g., those in Figs. \ref{fig_illustration} (a-c) (more details can be found in Sec. SII of \cite{Supp}).

Finally, we discuss how to realize the nonlocal surface measurement in Fig.~\ref{fig_surfaceT}(d) with the help of the existing nanotechnologies. We propose the following fabrication process flow. The bottom graphite electrodes can be prepatterned by electron beam lithography and plasma etching on Si/SiO$_2$ wafers. To ensure the contact, the polymer residues on the graphite electrodes need to be removed by forming gas annealing in Ar/H$_2$ environment. Thin flakes of MnBi$_2$Te$_4$ are mechanically exfoliated. Flakes with preferred thickness (e.g., 200 nm) and flatness can be selected by atomic force microscopy, then laminated to the prepatterned bottom electrodes via polymer-assisted dry transfer technique. Moreover, the quantum anomalous Hall effect has been reported in a MnBi$_2$Te$_4$/Bi$_2$Te$_3$ superlattice of a thickness up to 300 nm, in which the Fermi level is efficiently tuned in the surface gap~\cite{Deng2020NatPhy}. To electrically separate the bottom electrodes and avoid contact on the side edges of MnBi$_2$Te$_4$, lithography free contact in Ref.~\cite{Telford18nl} can be used. Such electrodes can be thin and flat so that it is non-invasive for probing just the surface septuple layer. The process involving MnBi$_2$Te$_4$ thin flakes has to be protected by an inert gas environment such as a glove-box, as surface reconstructions may occur~\cite{Hou20an}.
Therefore, all of these nanotechnologies have been there, offering the possibility of detecting the axion insulator.

\begin{acknowledgments}
We thank helpful discussions with Jianpeng Liu, Chuizhen Chen, Hua Jiang, Donghui Xu and Bin Zhou. This work was supported by the National Natural Science Foundation of China (11534001, 11974249, 11925402), the Strategic Priority Research Program of Chinese Academy of Sciences (Grant No. XDB28000000), Guangdong province (2016ZT06D348), the National Key R \& D Program (2016YFA0301700), the Natural Science Foundation of Shanghai (Grant No. 19ZR1437300), Shenzhen High-level Special Fund (No. G02206304, G02206404), and the Science, Technology and Innovation Commission of Shenzhen Municipality (ZDSYS20190902092905285, ZDSYS20170303165926217, JCYJ20170412152620376, KYTDPT20181011104202253). R.C. acknowledges support from the project funded by the China Postdoctoral Science Foundation (Grant No. 2019M661678) and the SUSTech Presidential Postdoctoral Fellowship. The numerical calculations were supported by Center for Computational Science and Engineering of Southern University of Science and Technology.
\end{acknowledgments}

\bibliographystyle{apsrev4-1-etal-title_6authors}
\bibliography{refs-transport}

\end{document}